\documentclass[12pt]{article}

\usepackage{cite}
\usepackage{tabularx}
\usepackage[nointegrals]{wasysym}
\usepackage{caption} 
\usepackage{fullpage}
\usepackage{amsmath,amssymb,amsfonts}
\usepackage{graphicx}
\usepackage{multirow}
\usepackage{float}
\usepackage{textcomp}
\usepackage{multirow}
\usepackage{booktabs}
\usepackage{algorithm}
\usepackage{algcompatible}
\usepackage[T1]{fontenc}
\def\BibTeX{{\rm B\kern-.05em{\sc i\kern-.025em b}\kern-.08em
    T\kern-.1667em\lower.7ex\hbox{E}\kern-.125emX}}
\begin{document}

\title{Discovery of Layered Software Architecture from Source Code Using Ego Networks\\}

\author{Sanjay Thakare$^1$, Arvind W Kiwelekar$^2$\\
Department of Computer Engineering \\
Dr. Babasaheb Ambedkar Technological University,\\Maharashtra, India.\\
1:mail2sbt@gmail.com,2:awk@dbatu.ac.in}

\maketitle

\begin{abstract}
Software architecture refers to the high-level abstraction of a system including the configuration of the involved elements and the interactions and relationships that exist between them. Source codes can be easily built by referring to the software architectures. However, the reverse process i.e. derivation of the software architecture from the source code is a challenging task. Further, such an architecture consists of multiple layers, and distributing the existing elements into these layers should be done accurately and efficiently. In this paper, a novel approach is presented for the recovery of layered architectures from Java-based software systems using the concept of ego networks. Ego networks have traditionally been used for social network analysis, but in this paper, they are modified in a particular way and tuned to suit the mentioned task. Specifically, a dependency network is extracted from the source code to create an ego network. The ego network is processed to create and optimize ego layers in a particular structure. These ego layers when integrated and optimized together give the final layered architecture. The proposed approach is evaluated in two ways: on static versions of three open-source software, and a continuously evolving software system. The distribution of nodes amongst the proposed layers and the committed violations are observed on both class level and package level. The proposed method is seen to outperform the existing standard approaches over multiple performance metrics. We also carry out the analysis of variation in the results concerning the change in the node selection strategy and the frequency. The empirical observations show promising signs for recovering software architecture layers from source codes using this technique and also extending it further to other languages and software.
\end{abstract}

\section{Introduction}
\par Software architecture has been defined as the set of components and connectors that satisfy predefined functional and quality attributes \cite{shaw1996software}. It is the most interpretable, significant, and irreplaceable part of the entire system design process. Software is built upon a certain architectural style. In most cases, the post-development phase consisting of maintenance, support, and evolution leads to changes in the system. The architectural documentation of the software may not be maintained with the same consistency and might not be up-to-date. Moreover, frequent changes in the involved classes and packages with the motivation of improvement in results would lead to deviation in the original underlying architecture. Thus, it is highly desired to have an architecture recovery algorithm that can reconstruct back and document the software architecture based on the latest software code configuration.
\par The most common type of software architecture format used in the development process is the layered architecture style. The layered architecture breaks down the system into different layers of abstraction and assigns every element to each of these layers based on their role and responsibility. Layered software architecture has lesser design constraints, is simplistic, and easy to construct. The layers are present in hierarchical order with specific rules governing the interactions between them. Lower layers are supposed to provide the required information to the top layers, while the top layers can only interact with the very next lower layers.
\par Based on these constraints on the interactions between layers, there are two types of layer styles: closed and open. In the closed layer style, any layer has access to the services of only the next lower layer. In contrast in an open layering style, multiple lower layers can communicate and provide services to a single top layer. Our work focuses on the recovery and validation of layered architectures from Java-based software systems that are closed layer in style.
\par Recovery of layered architectures has been an area of active interest in the software engineering community. Previously proposed approaches in this domain can be broadly classified into three categories: automatic, semi-automatic, and manual. Clustering \cite{andreopoulos2007clustering} is the most commonly used automatic technique and has also been coupled with other features \cite{belle2014recovering} for recovering the layers of Object-oriented software systems. Semi-automatic and manual techniques require the assistance of domain experts or the software developers, thus introducing extra costs and bias. The knowledge of a domain expert has been utilized previously \cite{sarkar2009discovery}. However, such resources may not be available with everyone and a superior fully automatic recovery method is highly desired.
\par In this paper, we present a novel automated approach to recover the layered software architecture for better interpretability of systems. The relations between the various components including classes and packages are parsed and extracted from Java-based software systems. A directed network is used to represent the extracted elements as nodes, while the edges are used to indicate the relationship between those elements. We consider this network as analogous to a social network and apply the concept of ego networks from social network analysis for recovering the layered software architecture.
\par Ego networks are a distinct kind of social network, quite similar to a mainstream network that consists of social units (people, organizations, etc.) connected by social relations \cite{freeman1982centered}. Ego networks are constructed uniquely. A specific unit from the network is considered as the focal point or the ego of the network. All the social relations are then defined with respect to this ego unit and other nodes. All the individuals that are directly connected with the ego unit are identified, and all such pairs are enlisted. These first connections form a mini-network that represents the social world from the ego nodes' perspective. Exchange of information takes place in both directions and is the most impactful one. This could be considered similar to people influencing the personality, attitude, and values of the ego.

\par Ego network is composed of immediate connections that are in the first degree to the ego node. A single link indicates the unit distance in the ego network. If the network size is extended by one, then the neighbors of neighbors are also included. In such a case, the length is set to two. This would mean there is an increase in the probability of the occurrence of unusual members \cite{newman2003ego}. In this paper, we consider the first-order or immediate neighborhood of the ego. We prefer to use this approach owing to the easiness in data collection, interpretability at a microscopic level, and better statistical inferences than those of the entire network.
\par Through this paper, our main contributions can be enlisted as follows:
\begin{itemize}
    \item A comprehensible literature review of the work done in this domain, with a thorough discussion on their advantages and shortcomings.
    \item A novel approach for software architecture recovery by the construction of ego network from the source code.
    \item Combination of various ego layers in a specific way for an optimized construction of the layered architecture.
    \item Demonstration of better performance in architecture recovery as compared to multiple previous approaches on three distinct software.
    \item Analysis of architecture recovery when operating on an evolving software system.
\end{itemize}

\par The outline of the rest of the paper is as follows:
Section 2 provides a detailed literature review of the previous work in this domain while section 3 describes the proposed architecture recovery approach. Section 4 provides information about the software systems on which the evaluation is done along with the obtained results. Section 5 discusses the validity and open issues in our proposed approach and Section 6 concludes this work.

\section{Background and Related work}

\par Several approaches have been proposed in the past for the retrieval of layered software architecture. Out of these approaches, clustering has been the most prevalent one. Most of the existing systems that are based on clustering have a two-phase recovery process. In the first phase, a set of classes are grouped into clusters based on defined criteria. This decomposes the system into sets of independent clusters. Next, these sets of clusters are used to form abstract layers. However, such an architecture is vulnerable to violations owing to its instability. Further, the process is not extensible to other software systems because each process requires a different set of inputs(structural, non-structural, or both) and are built only according to certain design specifications.
\par Laval et al. \cite{laval2013ozone} proposed a tool OZONE that detects the dependencies present in the software system and marks them as either removable or not-flagged. Removable dependencies are those that lead to the formation of either a direct or an indirect cycle. These predictions are further evaluated by a human expert who further classifies them into three categories of expected, undesirable, and removable. Removable and undesirable dependencies are those that lead to the breakage of the cycle and are ignored during the layer formation. Thus, the creation of an acyclic structure of the software simplifies the layer recovery process.
\par Heuristic-based approaches have also been proposed previously to tackle this problem. Andreopoulos et al. \cite{andreopoulos2007clustering} proposed heuristics that applied fan-out and fan-in dependencies of a module to place it to the top or bottom layer in the architecture. A system expert was used for further division of the remaining entities. Scanniello et al. \cite{scanniello2010architectural} calculated the authority and hub values for each node in the network and classified the nodes into the top and bottom layers. Higher authority and low hub classes occupy the top layer while the bottom layer consisted of classes with low authority and high hub values. A major shortcoming of these heuristic-based approaches is that none of them exploit the layering principles.
\par Some approaches have made use of resources like source code and design descriptions and have coupled them with information from the team or domain experts. Such approaches are termed semi-automatic and are more expensive in terms of both time and cost. The main benefit of such approaches is in cases where the documentation is not sufficient, interpretable, and up-to-date. Sarkar et al. \cite{sarkar2009discovery} and Haitzer and Zdun \cite{haitzer2015semi} had proposed such semi-automatic approaches for this task. Further, such information needed for grouping the modules into layers can be categorized into structured and unstructured information. Some layered architecture recovery approaches have utilized structure information like rank \cite{sarkar2010architecture}, fan-in, and fan-out of the projects \cite{salehie2005architectural} while other approaches have relied on unstructured information such as concept distribution \cite{sarkar2010architecture}. Sarkar et al. \cite{sarkar2009discovery} had combined both this information for better results.
\par Hussain et al. \cite{hussain2015novel} used evolutionary and genetic algorithms including Particle Swarm Optimization (PSO) for this task. While they achieved better results, the approaches relied excessively on the proper tuning of the parameters. Samrongsap and Vatanawood \cite{samrongsap2014tool} defined some architecture style rules that were used to execute the recovery based on a dependency syntax tree built using the Roslyn tool.
\par Belle et al. \cite{belle2013layered,belle2014recovering} proposed responsibility based clustering for layer recovery in object-oriented software systems. Package namespaces (fully qualified package names) were used for building the responsibility clusters. Barring the leaves, the package elements were those that demonstrate a high level of granularity and were used to create the responsibility tree of the system. Levels of abstraction were decided based on the granularity levels of the packages with all leaf packages being grouped into a single cluster. Optimization of the layer construction process from these clusters focused on the abstraction uniformity and incremental dependency. Abstraction uniformity refers to the assignment of the same layer to the clusters having the same level of abstraction. Incremental dependency expects the existence of dependency between the clusters present in the upper and the lower layer.
\par Several metrics have been proposed for evaluating the violations committed in the obtained layered architecture. Sarkar et al. \cite{sarkar2009discovery,sarkar2010architecture} proposed several indexes including back-call, skip-call, and cyclic violations for this purpose. In addition to these, coupling metrics were proposed by Saraiva et al. \cite{saraiva2010assessing,saraiva2011metrics} for measuring the coupling between the layers to follow the best practices for layered architecture. Salehie et al. \cite{salehie2005architectural} showed that classes with a higher number of children and variation in cohesion were more susceptible to faults and errors.

\par It can be seen that while significant work has been done in this domain, there is ample scope for improvement and amelioration of the shortcomings of previous techniques. We try to address these limitations in our proposed approach that is explained in the next section.

\section{Proposed methodology}
\par The problem that we are trying to solve is the accurate partitioning of the structural representation of the software system into distinct layers. Mathematically, the problem can be represented as follows: Consider a given network $G\ =\ \{V, E\}$ that is the structural representation of the software under consideration. Here, $V$ indicates the set of program entities and E denotes the set of links present between them. We need to partition $V$ into a distinct set of layers $L\ =\ \{l_1, l_2, l_3,..., l_n\}$ such that every entity in $V$ belongs to one and only one layers in $L$, while respecting the design constraints. As opposed to previous approaches working on observing common similarity patterns, we directly map the program elements into layers. We apply the concept of ego networks for this purpose. 
\par We define certain terminologies before diving deeper into explaining the methodology. The ego is the focal node of the personal/individual network. A personal network of the ego and its immediately connected neighbors is referred to as Ego network. The ego layer is a layer consisting of only an ego node. The ego layered structure is a three-layer structure where the middle layer always consists of the ego node. The nodes that connect to the ego node with an outgoing edge accommodate the top layer while the nodes connected by an incoming edge from the ego node are placed in the lower layer.

\begin{figure}[h]
    \centering
    \includegraphics[height=8cm]{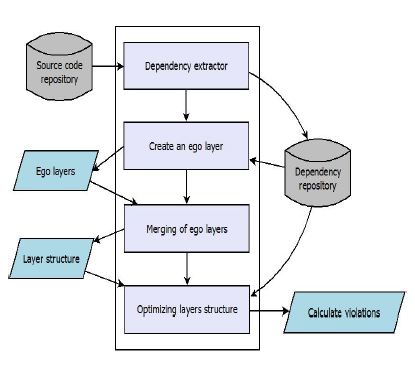}
    \caption{Block diagram of the proposed layer recovery approach}
    \label{block}
\end{figure}

\par The block diagram of the proposed layer recovery system using an ego network is shown in Figure \ref{block}. The first step consists of scanning the source code to retrieve the dependency information from the system under observation. The remaining steps are related to the ego layers extraction and reconstruction of the architecture. After the architecture is recovered, the layer violations are determined to assess the quality of architecture. 
\par We now elaborate upon the working of the algorithms involved in each of these phases in the next subsections.

\begin{figure}[h]
    \centering
    \includegraphics[height=7cm,width=\linewidth]{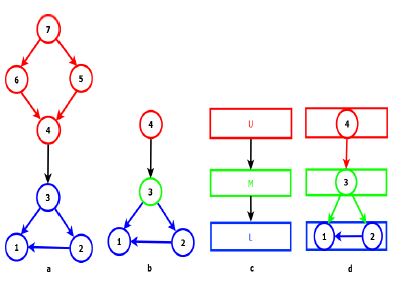}
    \caption{a) A sample dependency network b) An ego network created with node 3 as the ego node c) Empty three layer architecture, and d) Assigning ego network nodes to the layers}
    \label{create1}
\end{figure}

\subsection{Creation of an ego layer}
\par
\begin{figure}[h]
    \centering
    \includegraphics[height=7cm]{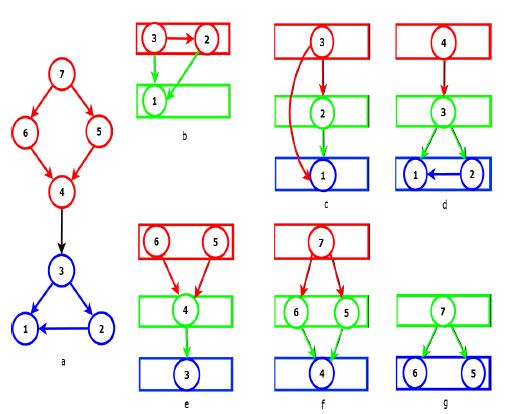}
    \caption{Creation of ego layer from a dependency network}
    \label{create2}
\end{figure}

 In this step, the dependency ego network is converted to the ego layers. A randomly selected ego node is assigned to an empty three-layer architecture in a way that the top layer accommodates nodes providing service to the ego node and the nodes in the bottom layers use the services of the ego node. The process of creating ego layers of the node involves three sub-processes. The steps followed for the creation of ego layer are depicted in Figure \ref{create1} and \ref{create2} and are explained below:
\begin{enumerate}
    \item Randomly select a node as an ego.
    \item Initialize a three ego layer structure for each ego network.
    \item Assign nodes of the ego network to each of the three layers based on the nature of their edge with respect to the ego node.
\end{enumerate}

\par Figure \ref{create2} represents the process of creating ego layers from the sample network 2a. Sub figure 3b represents an ego layer with node 1 as the ego node. The two incoming edges from node 2 and node 3 lead to those nodes being placed in the top layer and as there are no outgoing edges, the bottom layer is empty. Similarly, subfigures 3c, 3d, and 3e show the structure of the ego networks build with ego nodes set to nodes 2, 3, and 4 respectively. As node 5 and node 6 have similar incoming and outgoing connections, they share the same ego network structure shown in 3f. As node 7 does not have incoming connections, the ego network built around it has no top layer. To generalize, for a network with $n$ nodes, $n$ different ego networks can be created. For each of these ego networks, a three-layered structure is created. Thus, at most $3n$ layers are created for each initial software dependency network.

\begin{algorithm}[h]
 \caption{Creation of Ego layer} 
 \begin{algorithmic}[1]
 \renewcommand{\algorithmicrequire}{\textbf{Input:}}
 \renewcommand{\algorithmicensure}{\textbf{Output:}}
 \REQUIRE Dependency graph $G$ = \{Nodes $V [v_1,v_2,..,v_n]$,\ Edges\ $E$\}
 \ENSURE Ego Layer $eLayer$
 \FOR {$v_i$ in $V$}
 \STATE layer $\leftarrow$ createEmptyLayer()
 \STATE insertMiddleLayer(layer,$v_i$)
 \STATE insertDownLayer(layer,$G.getNeighbors(v_i)$)
 \FOR {$nbrnode$ in G.getNeighbors($v_i$)}
 \IF {isCreated($eLayer,nbrnode$)}
 \STATE insertUpLayerEle($layer, v_i,nbrnode$)
 \ELSE
 \STATE $layer2$ $\leftarrow$ createEmptyLayer()
 \STATE insertMiddleLayer($layer2, nbrnode$)
 \STATE insertUpLayerEle($layer, v_i, nbrnode$)
 \STATE insertLayer($eLayer,layer2$)
 \ENDIF
 \ENDFOR
 \STATE insertLayer($eLayer,layer$)
 \ENDFOR \\
 \textbf{return} $eLayer$
 \end{algorithmic}
 \end{algorithm}

\par Algorithm 1 shows the working of the proposed algorithm for the creation of ego networks. The runtime complexity of the proposed algorithm is $O(nk)$ where $n$ is the number of nodes and $k$ indicates the average degree of the network. It can be seen that $n$ iterations are always needed due to construction for each node in the network. However, the inner loop complexity depends on the number of neighbors with incoming edges to the ego node. This number would always be minimal as compared with $n$ and we consider it to be the average degree of the network.

\begin{figure*}
\includegraphics[width=\textwidth,height=11cm]{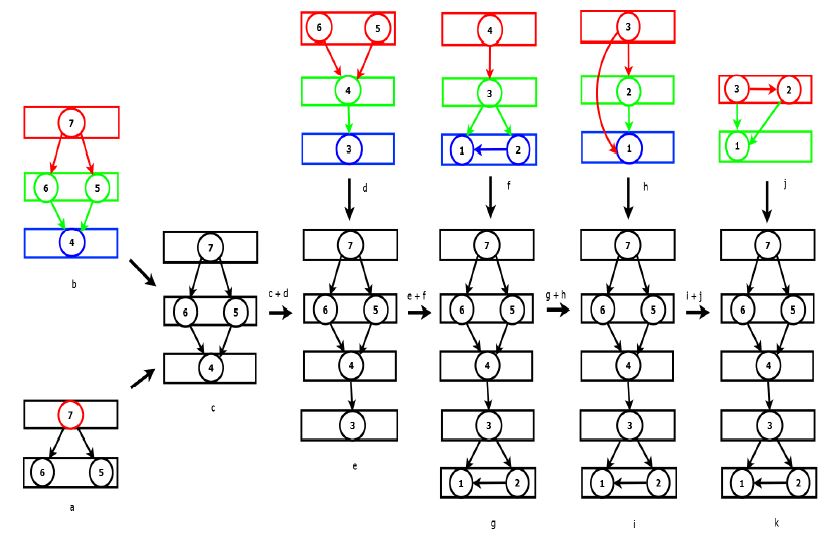}
\caption{Illustration of the layer merging process for various ego layers created from the sample dependency network}
\label{merge}
\end{figure*}

\subsection{Merging of ego layers}
\par After the creation of the ego network, merging is
the process of combining different layer structures into a single structure. This involves the identification and removal of duplicate layers. Duplicate layers are those that have similar nodes present as its members.
\par The merging process starts with the selection of the seed node that is done randomly. The selected node is referred to as ego, and its layered structure is extracted and initialized to derive the final result. From the remaining nodes, a node is selected at random. If it is present in the final layered structure, then its upper layer is merged with the top layer in the final layered structure and the bottom layer is combined with the bottom layer
of the final layered structure. Otherwise, the node is skipped and we move onto the next iteration. The process continues until all the layered structure of the nodes are not processed. This results in the creation of a layered structure that represents the current architecture of the software.
\par Figure \ref{merge} shows the merging process that takes the layered structure shown in Figure \ref{create2} as an input. Here, we start with node 7 as the seed node and its layered structure as the initial result as shown in Figure 4a. Then nodes 6 and 5 are selected and their layered structure (Figure 4b) is combined with layered structure 4a. First, node 5 or 6 are searched in the layered structure 4a. As they are present, the upper layer of node 5 or 6 that contains node 7 is merged with the top layer of 4a. Here, both the layer contains the same node, so there is no need to merge the layers. Otherwise, node 5 or 6 are not processed. The lower layer of nodes 5 and 6 consists of node 4 and are added to the newly created layer as shown in figure 4c. For the next iteration, node 4 is selected. It is also present in the partial structure. We ignore the upper layer of node 4 as it is similar to the partial structure and add the lower layer to structure as shown in figure 4e. In the subsequent iteration, as shown in figure 4g, the layer with node labeled 1 and 2 is added at the bottom of node 3. It can be seen that for the remaining iterations and nodes, the structure remains unchanged.
\par The detailed process of merging the layers is described in Algorithm 2. The runtime complexity of the proposed algorithm is $O(n)$ where $n$ is the number of nodes in the network. The complexity is majorly dependant on the implementation of the $searchNode$ function.

\begin{algorithm}[h]
 \caption{Merging of layers} 
 \begin{algorithmic}[1]
 \renewcommand{\algorithmicrequire}{\textbf{Input:}}
 \renewcommand{\algorithmicensure}{\textbf{Output:}}
 \REQUIRE Ego layer structure $eLayer$
 \ENSURE Merged layer $mLayer$
 \FOR {$node$ in $eLayer$}
 \STATE nel $\leftarrow$ getNodeEgoLayer($node$)
 \IF {isEmpty($mLayer$)}
 \STATE $mLayer$ = nel
 \ELSE
 \STATE $lay$ $\leftarrow$ searchNode($node$)
 \IF {valid lay}
 \STATE mergeUp($lay$, getUpLayer(nel))
 \STATE mergeDown($lay$, getDownLayer(nel))
 \ENDIF
 \ENDIF
 \ENDFOR \\
 \textbf{return} $mLayer$
 \end{algorithmic}
 \end{algorithm}

\subsection{Optimization of the extracted layer architecture}
\par In the merging process, a layered architecture is recovered from the dependency network of the software system. However, the layers present in this structure indicate the maximum possible layers that the software entities can be divided into. In Figure \ref{merge}, it can be seen that a five-layered structure is obtained as the output. The dependency depth of the sample network is also five. Hence, the maximum number of layers extracted in the worst case is equal to the dependency depth of the network. The dependency depth, also referred to as diameter is the longest chain of dependency in the network.
\par
\begin{figure}[h]
    \centering
    \includegraphics[height=5cm]{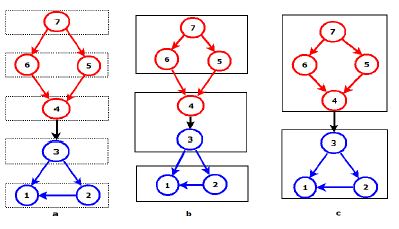}
    \caption{a) Initial Layered architecture created after merging the various ego layers of the network; b) and c) Optimized layer structure created from a)}
    \label{optimize}
\end{figure}

 In the optimization process, we remove the unnecessary layers of the extracted layered structure. The layers that are relatively smaller in size are merged with their adjacent layers. As shown in Figure \ref{optimize}, the original extracted structure consists of five layers. However, the top and the middle layers consist of only one node. The selection and merger of these layers with the adjacent ones lead to a layered structure similar to that shown in Figure 5b or 5c. Algorithm 3 elucidates the optimization of the layering structure. A threshold value is decided and layers with a size smaller than the threshold are combined with two adjacent layers, one present upward and another present on the downside.
\par The resultant structure is evaluated for its violation count:
\begin{equation}
    v\ =\ \alpha.B + \beta.S
\end{equation}
\par where $B$ and $S$ indicate the back-call and skip-call violations. $\alpha$ and $\beta$ are the weights assigned to the back-call and skip-call violations respectively. The structure that reduces the overall violations is retained, and hence layers are optimized. In instances where the adjacent layers are not present a direct combination of these layers is done. The runtime complexity of the proposed algorithm depends on the evaluation of the structure. During the evaluation process, we have to parse each edge present in the network to determine the back-call or skip-call violations. So, the evaluation process will take $O(E)$ time to measure the violations.

\begin{algorithm}[h]
 \caption{Optimization of layer structure} 
 \begin{algorithmic}[1]
 \renewcommand{\algorithmicrequire}{\textbf{Input:}}
 \renewcommand{\algorithmicensure}{\textbf{Output:}}
 \REQUIRE Merged layer structure $mLayer$, Threshold $\Delta$
 \ENSURE Optimized layer architecture $optLayer$
 \FOR {$layer$ in $mLayer$}
 \STATE c $\leftarrow$ getCurLayerNum($mLayer$)
 \IF {sizeOf($layer$) < $\Delta$}
 \STATE $merge1$ $\leftarrow$ combLayers($layer,getLayer(c-1)$)
 \STATE $v1$ $\leftarrow$ evaluate($merge1$)
 \STATE $merge2$ $\leftarrow$ combLayers($layer,getLayer(c+1)$)
 \STATE $v2$ $\leftarrow$ evaluate($merge2$)
 \IF {$v1 \leq v2$}
 \STATE insertLayer(c-1, merge1)
 \ELSE 
 \STATE insertLayer(c+1, merge2)
 \ENDIF
 \STATE deleteLayer(c)
 \ENDIF
 \ENDFOR \\
 \textbf{return} $optLayer$
 \end{algorithmic}
 \end{algorithm}

\section{Experimental results}
\par This section presents the experimental validation of our proposed approach to the selected software. The experiment consists of three parts, starting from the extraction of dependencies from the classes, then the recovery of the layered architecture, and finally the validation of the predicted architecture to check for the committed violations in the recovered layer organization. We present results for two types of software systems: static versions, and continuously evolving versions. For validation of static software, we have selected three software systems, two of which are libraries and one is an application. The analysis of evolving software is done on the JHotDraw software. Experimental validation in this domain is challenging owing to the lack of software systems that have both, up-to-date documentation and a mentioned layer structure.

\subsection{Evaluation on static software systems}
\subsubsection{Constore}
\par
\begin{figure}[h]
    \centering
    \includegraphics[height=6cm]{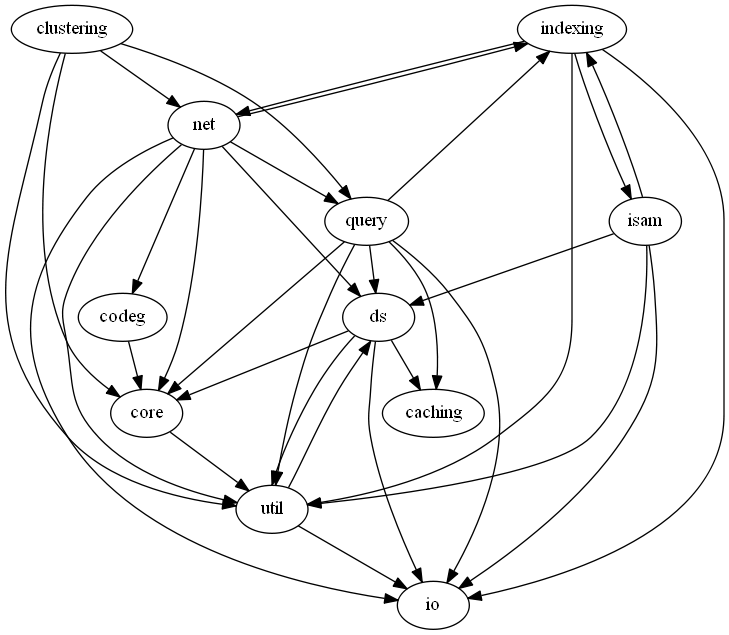}
    \caption{Package dependency of the Constore library}
    \label{constore}
\end{figure}

 ConStore is a small library written in Java that is used to model real-world problems using a concept network. A meta model holds the conceptual linking of the topic being considered. A detailed model is created based on the rules established in the meta model. Graphs are used to represent the models with the nodes indicating concepts and the edges indicating the relation between those concepts. A user interface has been provided to store, navigate and retrieve various concept networks \cite{constore}. Figure \ref{constore} indicates the package dependency graph of the library.

\begin{figure*} 
    \centering
    \includegraphics[height=8cm,width=\textwidth]{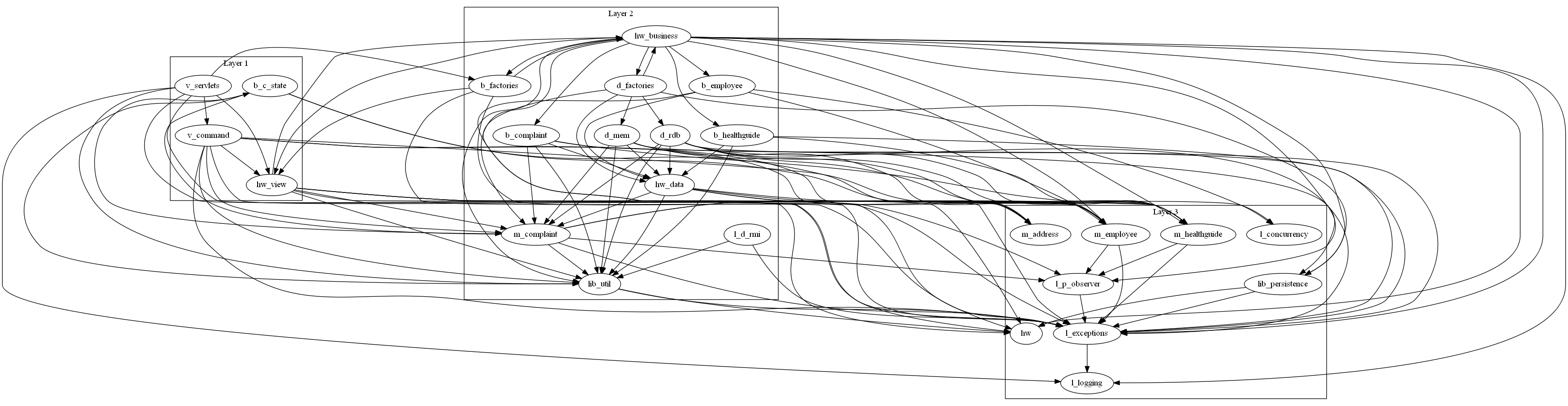}
    \caption{Layered architecture recovered from package dependency of the Health Watcher application using our proposed approach}
    \label{healthwatchla}
\end{figure*}

\begin{figure}[h]
    \centering
    \includegraphics[height=8cm,width=\linewidth]{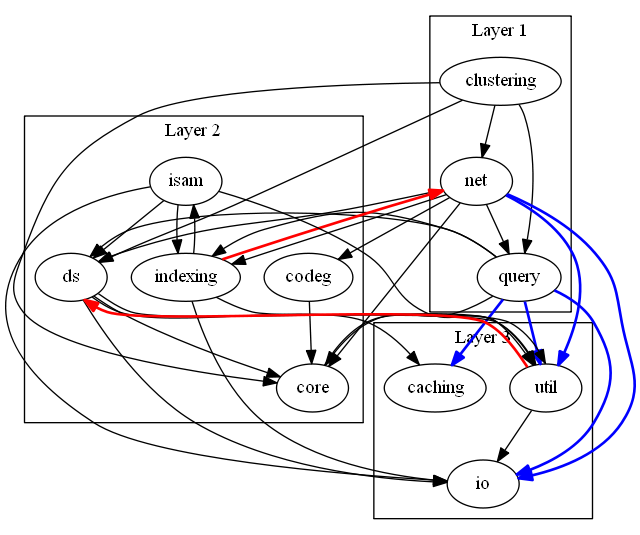}
    \caption{Layered architecture recovered from package dependency of the Constore library using the proposed approach}
    \label{constorelar}
\end{figure}

\par The overall back-call and skip-call violations for Constore at package level were recorded at 6\% and 13\% respectively. The skip-call violations are shown in Figure \ref{constorelar}, with \textit{blue} edges from the net and query packages to util and io packages. 
\par While we did detect five cycles in the Constore library, two of them were across layers. The length of the cycle responsible for the violation was three. The first cycle consisted of packages core, util, and ds, and the second cycle included the indexing, net, and query packages. In class level dependency analysis, we have recovered a similar architecture with more number of layers. The overall back-call and skip-call violations for Constore at the class level were found to be 13\% and 10\%. After optimization of the layers, the average back-call and skip-call violations dropped to up to 10\% and 3\% respectively.

\subsubsection{Health Watcher\\}
\par Health Watcher is a web-based application for addressing health-related problems \cite{greenwood2007impact}. A user can register, update, and query their health-related complaints and problems. The application follows a layered architecture style and the related information is known in advance. It implements an open layering style wherein a library is shared amongst all the layers.
\par The overall back-call and skip-call violations for the Health Watcher application at package level were noted at 3\% and 14\%. The extracted layered architecture is shown in Figure \ref{healthwatchla}. Five cycles have been detected, one of them having a length three and being responsible for the layer violations. The number of skip-calls violations for this software can be attributed to the distribution of library package amongst the layers and most of the libraries being defined in the lower layers where they are supposed to be present. The overall back-call and skip-call violations for Health Watcher Application at the class level are 3\% and 8\%. Optimization of the layers led to a further reduction in skip-call violations to 5\%.

\subsubsection{JFreeChart}
\par JFreeChart is a Java-based library extensively used for drawing charts embedded in various applications and web pages \cite{gilbert2007jfreechart}. In our evaluation, back-call in the JFreeChart library was observed mostly between the middle and the upper layers of the architecture. Only a single back-call was found between the top two layers. 
\par The overall back-call and skip-call violations in JFreeChart were recorded at 0.004\% and 20\% respectively. 
Many cycles (around 3176) were observed, with most of them spanning beyond a single layer, more often reaching to all of the layers. A heavy package dependency could potentially impact the overall maintainability of the system.


\begin{table}
 \centering
 \begin{tabular}{|p{2.5in}|p{0.6in}|p{0.6in}|p{0.6in}|p{0.6in}|}
 \hline
 Software Methods & Recall & Precision & F-Score & Accuracy \\ \hline 
  \multicolumn{5}{|p{4.9in}|}{\centering{\bf Constore}} \\ \hline 
  BBA\cite{belle2014recovering} & 0.44 & 0.44 & 0.42 & 45.45 \\ \hline
Ozone\cite{laval2013ozone} & 0.44 & 0.44 & 0.42 & 81.82  \\ \hline
HubAuth\cite{scanniello2010architectural} & 0.44 & 0.44 & 0.42 & 69.23  \\ \hline
Bunch\cite{mancoridis1999bunch} & 0.64 & 0.64 & 0.63 & 63.64  \\ \hline
  
 \textbf{Ego network (Proposed)} & \textbf{0.92} & \textbf{0.93} & \textbf{0.92} & \textbf{90.91}   \\ \hline
 
   \multicolumn{5}{|p{4.9in}|}{\centering{\bf HW}} \\ \hline 
  
  BBA\cite{belle2014recovering}  & 0.70 & 0.59 & 0.59 & 62.07  \\ \hline
Ozone\cite{laval2013ozone}   & 0.75 & 0.68 & 0.65 & 62.07   \\ \hline
HubAuth\cite{scanniello2010architectural}   & 0.83 & 0.81 & 0.81 & 86.21  \\ \hline
Bunch\cite{mancoridis1999bunch}   & 0.77 & 0.67 & 0.66 & 65.52   \\ \hline
\textbf{Ego network (Proposed)}   & \textbf{0.92} & \textbf{0.83} & \textbf{0.86} & \textbf{86.21}   \\ \hline
  
  \multicolumn{5}{|p{4.9in}|}{\centering{\bf JFreeChart}} \\ \hline 
  
  BBA\cite{belle2014recovering} &   0.64 & 0.54 & 0.53 & 51.16 \\ \hline
Ozone\cite{laval2013ozone}   & 0.59 & 0.54 & 0.42 & 41.86 \\ \hline
HubAuth\cite{scanniello2010architectural} &   0.56 & 0.56 & 0.55 & 60.47 \\ \hline
Bunch\cite{mancoridis1999bunch} &   0.53 & 0.46 & 0.44 & 44.19 \\ \hline
\textbf{Ego network (Proposed)} &   \textbf{0.86} & \textbf{0.81} & \textbf{0.81} & \textbf{81.40} \\ \hline

 \end{tabular}
\caption{Comparison with standard architecture recovery approaches on multiple performance metrics}
\label{tab:tab_stat_precisionrecall}
\end{table}

\begin{table}[h]
\centering
\renewcommand{\arraystretch}{1}
\begin{tabular}{|p{0.1\linewidth}|p{0.07\linewidth}|p{0.07\linewidth}|p{0.05\linewidth}|p{0.07\linewidth}|p{0.07\linewidth}|p{0.05\linewidth}|p{0.15\linewidth}|}
\hline
\multicolumn{2}{|l|}{\textbf{\begin{tabular}[c]{@{}l@{}}Approach\end{tabular}}} & Actual & BBA   & Ozone       & \begin{tabular}[c]{@{}l@{}}Hub-\\ Auth\end{tabular} & Bunch & \textbf{\begin{tabular}[c]{@{}l@{}}Ego net\\ (Proposed)\end{tabular}} \\ \hline
\multirow{3}{*}{Constore}                                                      & Back       & 2      & 15    & 3           & 0                                                   & 7     & \textbf{2}                                                            \\ \cline{2-8} 
                                                                               & Skip       & 5      & 1     & 8           & 0                                                   & 6     & \textbf{5}                                                            \\ \cline{2-8} 
                                                                               & Cyclic     & 2      & 2     & 2           & 2                                                   & 2     & \textbf{2}                                                            \\ \hline
\multirow{3}{*}{\begin{tabular}[c]{@{}l@{}}Health-\\ Watcher\end{tabular}}     & Back       & 2      & 1     & 3           & 0                                                   & 9     & 3                                                                     \\ \cline{2-8} 
                                                                               & Skip       & 17     & 25    & 6           & 12                                                  & 9     & \textbf{14}                                                           \\ \cline{2-8} 
                                                                               & Cyclic     & 1      & 1     & 1           & 1                                                   & 1     & 1                                                                     \\ \hline
\multirow{3}{*}{\begin{tabular}[c]{@{}l@{}}JFree-\\ Chart\end{tabular}}        & Back       & 12     & 0     & \textbf{12} & 0                                                   & 29    & 1                                                                     \\ \cline{2-8} 
                                                                               & Skip       & 33     & 51    & \textbf{33} & 6                                                   & 30    & 51                                                                    \\ \cline{2-8} 
                                                                               & Cyclic     & large  & large & 0           & large                                               & large & large                                                                 \\ \hline
\end{tabular}
\caption{Analysis of the committed layer violations}
\label{tab_stat_vio}
\end{table}

\par We compare the performance of our proposed approach with four other standard approaches. These approaches are as follows:
\begin{itemize}
    \item Bunch \cite{mancoridis1999bunch}: Bunch is a clustering-based tool that focuses on the recovery of the layered architecture using the best possible partitions amongst the software elements based on the similarity between them.
    \item Scaniello et al. \cite{scanniello2010architectural}: A static link analysis approach that identifies the relationship between the classes present in the system and then uses link analysis to group them into layers.
    \item Laval et al. \cite{laval2013ozone}: The proposed tool makes use of a heuristic-based approach to detect and work upon the cyclic dependencies present in the software network to derive the architecture.
    \item Belle et al. \cite{belle2014recovering}: The approach focused on the use of clustering, followed by optimization and derivation of granularity levels to find the recovered clusters.
\end{itemize}

\par We evaluate the correctness of the recovered architecture in two ways:
\begin{enumerate}
    \item Violation of layering principles \cite{sarkar2010architecture}
    \item Standard classification metrics: Precision, recall, F-score, and accuracy
\end{enumerate}

\par Table \ref{tab_stat_vio} summarizes the violations in layering committed by applying each of the mentioned approaches and our proposed approach. The second column mentions the actual layering of the software systems. The back-call, skip-call, and cyclic violation values are reported for each of the software systems.

\par The obtained result indicates that the violations of the extracted architecture by applying the proposed network closely match with the actual layered architecture for Constore and HW application, whereas for JFreechart, the results deviate from the ground truth. The main reason for this deviation is due to the huge number of cyclic dependency between the packages. The Ozone approach recovered the true architecture for JFreechart because the approach first removes the cyclic dependencies and then layers the acyclic dependency network. In our approach, we recovered the architecture in the presence of cyclic dependency.

\par Table \ref{tab:tab_stat_precisionrecall} summarizes the results obtained in terms of the standard performance metrics wiz. accuracy, precision, recall, and F1 score. The obtained results show that our approach had better accuracy, thus implying that better identification of layers has been done. Also, the F-score values of our approach are much better than the remaining approach for all the software systems. Further, the precision and recall values of our approach show that the approach would correctly and completely recover most of the software layers. Generally, we can say that the approach is correct and complete for all the software systems because the precision values are slightly greater than the recall values for all the considered software systems. Thus an all-round evaluation across multiple metrics has been done to demonstrate the advantage of our proposed approach. In the future, we intend to analyze the results of even larger software systems.

\begin{table}[h]
\centering
\renewcommand{\arraystretch}{1.2}
\begin{tabular}{|l|l|l|l|l|l|l|l|l|l|}
\hline
\multirow{2}{*}{\textbf{\begin{tabular}[c]{@{}l@{}}Software\\ method\end{tabular}}} & \multicolumn{3}{l|}{\textbf{Constore}} & \multicolumn{3}{c|}{\textbf{HW}} & \multicolumn{3}{l|}{\textbf{JFreechart}} \\ \cline{2-10} 
                                                                                    & B           & M           & T          & B         & M         & T        & B            & M           & T           \\ \hline
Actual  & 3 & 4 & 4 & 17 & 9  & 3 & 13 & 23 & 7  \\\hline
BBA     & 3 & 6 & 2 & 11 & 10 & 8 & 12 & 12 & 19 \\\hline
Ozon    & 3 & 4 & 4 & 8  & 16 & 5 & 13 & 3  & 27 \\\hline
HubAuth & 2 & 8 & 1 & 14 & 12 & 3 & 16 & 22 & 5  \\\hline
Bunch   & 4 & 3 & 4 & 9  & 14 & 6 & 16 & 10 & 17 \\\hline
\textbf{\begin{tabular}[c]{@{}l@{}}Ego networks\\ (proposed)\end{tabular}}    & 3 & 5 & 3 & 13 & 12 & 4 & 11 & 19 & 13 \\\hline
\end{tabular}
\caption{Distribution of nodes in the three-tier architecture}
\label{tab:tab_stat_dist}
\end{table}

\par Table \ref{tab:tab_stat_dist} summaries the number of nodes present in each identified layer (T=Top, M=Middle, B=Bottom). The distribution of the nodes in the identified layers of our approach is nearly equal to the actual distribution.


\subsection{Analysis of software system evolution }
\par The previously mentioned results were for static versions of the software. To further prove the effectiveness of our proposed approach, we assess the performance for an evolving version of the JHotDraw software. The evolution is done from version 5.1 to version 7.0.6. Specifically, we have recovered the layers in two ways:
\begin{itemize}
    \item Separately reapplying the ego network approach for each version of the software
    \item Incremental approach consisting of extraction of layered architecture from the basic version and then updation of new information from the next version to the same.
\end{itemize}
\par The incremental layered extraction is based on the incremental clustering methodology proposed by (Fazli Can can1993incremental). In the case of incremental layering, we have used two seed architectures: version 5.1 due to it being the first release, and version 7.0.6 because the software was re-modernized by adding $63$ new packages and only retaining $3$ packages from the old architecture. For version 7.0.6, it is apparent that the entire structure of the software is changed completely due to almost 95\% packages being changed.
\par Our work differs in multiple dimensions from the research carried out by (Scanniello 2010): 1) We have analyzed package dependency network rather than the class-level dependency network, 2) We have considered more versions ($15$) of the JHotDraw software(from initial to the latest version), 3) The application of the concept of the ego networks and 4) the use of incremental layering. Further, we have assessed the system by both \textit{MoJo} distance and \textit{Stability}.

\par The tables \ref{tab:tab_evol_dist_jhd} and \ref{tab:tab_evol_stab_jhd} summarize the result of ego network approach to understand the evolution of JHotDraw software. In particular, Table \ref{tab:tab_evol_dist_jhd} provides information about the traditional and incremental layer recovery using the ego networks. Apart from the serial number, version of the software, and the number of packages in the respective version, the remaining columns indicate the distribution of the number of packages under traditional and incremental approaches. Specifically, the distribution is different for each version in the traditional approach irrespective of the changes. However, the incremental approach revealed similar distribution when no change was made in the software version and showed different distribution when new packages were added into the software.
\par Therefore, the first three versions under the incremental approach had similar distribution as there is no change in the software. It is observed that the top layer is almost unchanged in most of the later versions while the middle layer expands due to the addition of new nodes.

\begin{table}[h]
\centering
\begin{tabular}{|l|l|l|l|l|l|l|l|l|}
\hline
\multirow{2}{*}{\begin{tabular}[c]{@{}l@{}}Sr.\\ No.\end{tabular}} & \multirow{2}{*}{Version} & \multirow{2}{*}{Pack} & \multicolumn{3}{l|}{Traditional} & \multicolumn{3}{l|}{Incremental} \\ \cline{4-9} 
                                                                   &                          &                       & B         & M         & T        & B         & M         & T        \\ \hline
1  & jhd5\_1     & 10  & 3  & 4  & 3  & 3  & 3  & 4  \\\hline
2  & jhd5\_2     & 10  & 3  & 5  & 2  & 3  & 3  & 4  \\\hline
3  & jhd5\_3     & 10  & 3  & 5  & 2  & 3  & 3  & 4  \\\hline
4  & jhd5\_4b1   & 29  & 5  & 19 & 5  & 7  & 12 & 10 \\\hline
5  & jhd5\_4b2   & 29  & 4  & 19 & 6  & 7  & 12 & 10 \\\hline
6  & jhd6\_0\_b1 & 29  & 4  & 19 & 6  & 7  & 12 & 10 \\\hline
7  & jhd7\_0\_6  & 66  & 26 & 28 & 12 & 25 & 28 & 13 \\\hline
8  & jhd7\_0\_8  & 58  & 24 & 21 & 13 & 18 & 26 & 14 \\\hline
9  & jhd7\_0\_9  & 92  & 35 & 35 & 22 & 34 & 39 & 19 \\\hline
10 & jhd7\_1     & 81  & 21 & 18 & 42 & 32 & 34 & 15 \\\hline
11 & jhd7\_2     & 94  & 31 & 41 & 22 & 37 & 42 & 15 \\\hline
12 & jhd7\_3\_1  & 95  & 29 & 47 & 18 & 25 & 42 & 18 \\\hline
13 & jhd7\_4\_1  & 100 & 27 & 38 & 35 & 28 & 55 & 17 \\\hline
14 & jhd7\_5\_1  & 103 & 38 & 32 & 33 & 31 & 55 & 17 \\\hline
15 & jhd7\_6     & 107 & 18 & 49 & 40 & 30 & 60 & 17\\\hline
\end{tabular}
\caption{Distribution of nodes for evolving versions of JHotDraw}
\label{tab:tab_evol_dist_jhd}
\end{table}

\par The evolving layered architecture of the JHotDraw software is assessed using the MoJo distance and stability. It is used to determine the impact of changes on the erosion of the architecture. We have used MoJo (Move Operation and Join Operation) distance measure developed by Tzerpos and Holt to determine the distance between the two consecutive versions of the JHotDraw software system. The distance value is defined as the minimum number of moves and join operations that are required to be performed to transform one structure into another. During move operation, a package is shifted from one structure to another structure, while a join operation merges them into a single structure. In addition to this, a MoJoFM(FM) effectiveness measure express the distance in percentage.
\par The second measure for assessing the architecture of an evolving software system is stability. Stability is an indicator of the reaction of the system towards any applied changes. Less stable systems are fragile and could easily break due to even a small change. They have poor maintainability and any change has to be accompanied by rigorous testing and validation. The probability of occurrence of an error is also high. On the other hand, more stable systems are robust, accommodate local changes, and are more interpretable in nature. The stability value is derived by observing the impact and changes in the system occurring due to modifications for every element. The individual stability numbers are then averaged out to find the overall stability number in the system.
\par We first compute the Average Impact from the dependency network. The mathematical equation for which is inspired by the method proposed by \cite{stability2015}.
\begin{equation}
V_{impact} = \frac{1}{k} \sum_{i=1}^{k} \frac{1}{d}
\end{equation}
\begin{equation}
    I_{avg} = \frac{1}{N} \sum_{i=1}^{N} V_i
\end{equation}
\begin{equation}
    stability = (1 - I_{avg}) * 100
\end{equation}
Here, $d$ stands for the depth or level of impact and starts from 1. It indicates that the immediately connected nodes from the existing structure have a direct impact represented by 1. In the next level, the neighbors of neighbors have a $d$ value of 2. The average impact of a node is the average sum of impact received from each outgoing connection. The number of outgoing connections of a node is represented by $k$.
Stability and percentage Average Impact are opposite to each other. A system having stability of 60\% would mean that almost 40\% of the system elements are affected when any element is changed. It should be noted that stability is indicated as a ratio, and thus any calculations and subsequent analysis of the system is independent of the scale or size of the system.
\par Table \ref{tab:tab_evol_stab_jhd} summarizes the stability and Mojo distance for 15 versions of the JHotDraw software system. The Mojo distance is stable up to version 7.0.6 but gradually increases later due to an increase in the complexity of the system. The Mojo distance considers only the common element between two consecutive versions and their layering. The last two columns in the table represent the impact and stability of the version. In some cases, more changes have led to less impact while in other cases the introduction of lesser changes has produced a higher impact. The impact variation is based on the connection of new packages to the existing package structure. The stability of the latest version of JHotDraw is approximately 45\%, hence indicating that the recent system has now become more sensitive to the changes. 

\begin{table}[h]
\centering
\begin{tabular}{|p{0.1\linewidth}|l|p{0.05\linewidth}|p{0.05\linewidth}|l|l|p{0.03\linewidth}|l|l|}
\hline
\multirow{2}{*}{Version} & \multicolumn{4}{l|}{Packages} & \multicolumn{2}{l|}{Mojo} & \multicolumn{2}{l|}{Stability} \\ \cline{2-9} 
                         & Old    & New   & All   & \%   & Fm           & B          & Imp           & Stab           \\ \hline
jhd5\_1     & 10  & 0   & 10    & 0        & -     & -  & -      & -         \\ \hline
jhd5\_2     & 10  & 0   & 10    & 0        & 60    & 2  & 0      & 100       \\\hline
jhd5\_3     & 10  & 0   & 10    & 0        & 100   & 0  & 0      & 100       \\\hline
jhd5\_4b1   & 10  & 19  & 29    & 65.52    & 100   & 0  & 32.88  & 67.13     \\\hline
jhd5\_4b2   & 29  & 0   & 29    & 0        & 95.24 & 1  & 0      & 100       \\\hline
jhd6\_0\_b1 & 29  & 0   & 29    & 0        & 100   & 0  & 0      & 100       \\\hline
jhd7\_0\_6  & 3   & 63  & 66    & 95.45    & 95.24 & 1  & 100    & 0         \\\hline
jhd7\_0\_8  & 49  & 9   & 58    & 15.52    & 60    & 9  & 50.28  & 49.72     \\\hline
jhd7\_0\_9  & 57  & 35  & 92    & 38.04    & 80.95 & 4  & 36.90  & 63.10     \\\hline
jhd7\_1     & 77  & 4   & 81    & 4.94     & 77.5  & 9  & 10.12  & 89.88     \\\hline
jhd7\_2     & 81  & 13  & 94    & 13.83    & 62.26 & 15 & 72.18  & 27.82     \\\hline
jhd7\_3\_1  & 93  & 2   & 95    & 2.11     & 85    & 9  & 0      & 100       \\\hline
jhd7\_4\_1  & 83  & 17  & 100   & 17       & 94.12 & 3  & 33.46  & 66.54     \\\hline
jhd7\_5\_1  & 100 & 3   & 103   & 2.91     & 55.36 & 25 & 56.24  & 43.76     \\\hline
jhd7\_6     & 102 & 5   & 107   & 4.67     & 71.43 & 18 & 55.54  & 44.46    \\ \hline
\end{tabular}
\caption{Stability analysis of the JHotDraw software}
\label{tab:tab_evol_stab_jhd}
\end{table}

\section{Validity and open issues}
\par In this section, we present the threats posed to the validity of our results considering the previous studies in this domain:
\begin{itemize}
    \item Construct validity: Due to the limitations of high cost and time required, a threat could be posed upon the size of the experiments performed. While three software systems of varying sizes and types have been studied and evaluated, a larger number may be required to derive a more generalized conclusion.
    \item Internal validity: There is no presence of causality in our dependency graph and as a result, there is no threat to the internal validity of our approach.
    \item External validity: Our work focused on closed-layer systems that are Java-based and object-oriented in nature. The analysis was performed on the package level and class level. The work must be extended further to other object-oriented languages and other programming paradigms to avoid external validity threats.
    \item Reliability: A thorough and disinterested analysis of the back-call and skip-call violations was performed, thus reducing the possibility of any human evaluation error. However, there may be an introduction of error owing to any faulty dependencies extracted by the dependency extractor module.
\end{itemize}
\par Apart from these possible threats, we present the community with certain open issues that they could work upon to further improve the performance of the system:
\begin{enumerate}
    \item To check if the proposed network approach can generalize across other software types and methods.
    \item To come up with a relation between the cyclic dependency amongst the extracted architectures and the violation results.
    \item To explore the use of such an approach to non-object-oriented systems.
    \item To check for any correlation between the node ordering and the derived output during the layer merging process
\end{enumerate}

\section{Conclusion}
\par In this paper, a novel approach was proposed for the recovery of layered software architecture from the source code by deriving inspiration from the working of ego networks in social network analysis. The proposed approach ameliorated the shortcomings of the previous approaches, deviated from the conventional method of clustering, and performed a fully automated recovery of the architecture. We represented the dependencies in the software system as an ego network and then used this ego network for subsequent extraction and optimization of layers. The work was carried out on both static software and dynamically updating software. Experiments were conducted on multiple software and the performance was compared against four standard approaches. The results revealed that not only did our approach extract the layers accurately but also correctly identified the layers for the program elements, thus proving to be helpful for further documentation and maintenance. Future work for our proposed approach includes evaluation of the approach on heavy and costlier complex software systems along with the application of the method on various legacy and enterprise software systems. The node selection approach could also be worked upon to introduce some order instead of having a random selection approach in various phases of the extraction system. Our proposed approach has shown promising signs for the extension of this method to obtain better results across multiple disciplines in an effective yet efficient manner.

\bibliographystyle{plain}
\bibliography{ego_bib}

\end{document}